\documentclass[runningheads]{llncs}
\usepackage{graphicx}
\usepackage{multirow}
\usepackage[tikz]{ocgx2}
\usepackage{pgfplots}
\pgfplotsset{compat=1.18}
\usepackage{pgfplotstable}
\usepackage{pgf-pie}
\usepackage{subcaption}
\usepackage[title]{appendix}
\usepackage{wasysym}
\usepackage{enumitem}
\usepackage{color}
\usepackage{forloop}
\usepackage{ifthen}
\newcommand{\Qq}[1]{\textbf{#1}}

\newcounter{qr}

\newcommand{\Qline}[1]{\noindent\rule{#1}{0.6pt}}

\newcounter{ql}

\newenvironment{Qlist}{%

\begin{itemize}[leftmargin=1.5em,topsep=-.5em]
}{%
\end{itemize}
}

\newlength{\qt}

\newcounter{itemnummer}
\newcommand{\Qitem}[2][]{
\ifthenelse{\equal{#1}{}}{\stepcounter{itemnummer}}{}
\ifthenelse{\equal{#1}{a}}{\stepcounter{itemnummer}}{}
\begin{enumerate}[topsep=2pt,leftmargin=2.8em]
\item[\textbf{\arabic{itemnummer}#1.}] #2
\end{enumerate}
}

\definecolor{bgodd}{rgb}{0.8,0.8,0.8}
\definecolor{bgeven}{rgb}{0.9,0.9,0.9}
\newcounter{itemoddeven}
\newlength{\gb}
\newcommand{\QItem}[2][]{
\setlength{\gb}{\linewidth}
\addtolength{\gb}{-5.25pt}
\ifthenelse{\equal{\value{itemoddeven}}{0}}{%
\noindent\colorbox{bgeven}{\hskip-3pt\begin{minipage}{\gb}\Qitem[#1]{#2}\end{minipage}}%
\stepcounter{itemoddeven}%
}{%
\noindent\colorbox{bgodd}{\hskip-3pt\begin{minipage}{\gb}\Qitem[#1]{#2}\end{minipage}}%
\setcounter{itemoddeven}{0}%
}
}

\begin{document}
\title{User Perception of CAPTCHAs: A Comparative Study between University and Internet Users}
%
%
\author{Arun Reddy\inst{1} \and
Yuan Cheng\inst{2}\orcidID{0000-0001-7176-3951} 
}
\authorrunning{A. Reddy et al.}
%
\institute{
Intel Corporation, Santa Clara CA 95054, USA\\
\email{r3ddy.arun@gmail.com} \and
Grand Valley State University, Allendale MI 49401, USA\\
\email{chengy@gvsu.edu}}
\maketitle              
\begin{abstract}
CAPTCHAs are commonly used to distinguish between human and bot users on the web. 
However, despite having various types of CAPTCHAs, there are still concerns about their security and usability. 
To address these concerns, we surveyed over 250 participants from a university campus and Amazon Mechanical Turk. 
Our goal was to gather user perceptions regarding the security and usability of current CAPTCHA implementations. 
After analyzing the data using statistical and thematic methods, we found that users struggle to navigate current CAPTCHA challenges due to increasing difficulty levels. 
As a result, they experience frustration, which negatively impacts their user experience. 
Additionally, participants expressed concerns about the reliability and security of these systems. 
Our findings can offer valuable insights for creating more secure and user-friendly CAPTCHA technologies.

\end{abstract}
\section{Introduction}
The ``Completely Automated Public Turing Test to tell Computers and Humans Apart'' (CAPTCHA) \cite{von2004telling} is a popular challenge-response-based test that distinguishes real human users from bots on websites. 
CAPTCHAs prevent abuse such as false form submissions, fraudulent purchases, spam emails, and fake registrations.
Since its inception in the late 1990s, CAPTCHAs have taken on various forms, including text, image, audio, and video. 
However, security concerns have arisen as CAPTCHAs have become more commonplace on major websites.   
Text-based CAPTCHAs are vulnerable to optical character recognition (OCR) technology \cite{nguyen2014security,mori2003recognizing,yan2008low}, while image-based CAPTCHAs are susceptible to machine learning attacks \cite{golle2008machine}. 
Meanwhile, the security of audio and visual CAPTCHAs can be compromised by Hidden Markov Model-based attacks \cite{sano2015hmm}.
Even other types of CAPTCHAs have been found to be vulnerable under the influence of cookie tracking, tag frequency-based attacks, side-channel attacks, and simple JavaScript code \cite{sivakorn2016m,kluever2009balancing,hernandez2010pitfalls,vitas2019bypass}.

An ideal CAPTCHA should ace both security and usability. 
However, to defend against attacks, CAPTCHA systems often introduce distortion and noise to their challenges, making them more difficult for users to solve. 
Text-based CAPTCHAs, in particular, often contain scattered lines, dots, and distorted characters that require extra user effort.  
Furthermore, these distortions can be especially unfriendly to users whose native language does not use the Latin alphabet \cite{yan2008usability}.
Image-based CAPTCHAs can also present accessibility challenges for users with visual impairments or color blindness \cite{yan2008usability}.
On the other hand, audio and video-based CAPTCHAs may have issues with large file sizes and limited time for users to comprehend the content \cite{bursztein2010good,sauer2008towards}. 

Creating a secure yet user-friendly CAPTCHA system poses complex challenges. 
To gain insight into what this type of system should entail, we set out to answer three research questions:
\begin{itemize}
    \item What do users prefer in CAPTCHAs?
    \item How usable are current CAPTCHAs?
    \item How can we effectively balance security and usability when implementing CAPTCHAs?
\end{itemize}

We surveyed over 250 participants from two groups, including computer science students at a public university and individuals from Amazon Mechanical Turk. 
The survey evaluated the user experience of CAPTCHAs through quantitative and qualitative methods. 
We found that illegible texts/images and unclear instructions were the main reasons for failing a CAPTCHA challenge in one attempt. 
Additionally, human errors like lack of time and inattention also contributed to failures. 
We discovered that when users could not solve a CAPTCHA in one attempt, they often abandoned webpages, causing a decrease in traffic. 
The study revealed that users do not enjoy solving CAPTCHAs but see it as a necessary burden. 
We also analyzed the differences in responses between the two user groups and explored factors contributing to these variances.

In the following sections, we review relevant literature in Section \ref{sec:related}. 
We then detail the user study design in Section \ref{sec:method}.
In Section \ref{sec:findings}, we present the quantitative and qualitative findings.
We discuss the findings and suggest future work in Section \ref{sec:discussions} before the concluding remarks in Section \ref{sec:conclusion}.

\section{Related Work}\label{sec:related}
In this section, we will discuss the literature relevant to our study.

\subsection{CAPTCHA Security}
Chellapilla et al. argued that the difficulty of a CAPTCHA is ideal if humans solve the challenge with a success rate of at least 90\% while automatic scripts achieve a success rate of less than 0.01\% \cite{chellapilla2005designing}. Below, we outline some existing techniques for breaking common CAPTCHA types, including text-based, image-based, and audio/video-based.

Text-based CAPTCHAs, typically using alphanumeric characters, are vulnerable to optical character recognition (OCR) algorithms.
This technique breaks down the image into smaller segments containing a single character. It then matches against standard letter templates using a pattern recognition algorithm. 
Yan et al. presented a low-cost attack on text-based CAPTCHAs developed by tech giants such as Microsoft, Google, and Yahoo \cite{yan2008low}. The authors developed an attack that could segment the challenges generated. It also identified the exact location of all the characters in the order they appeared. 
Nguyen et al. discovered a method to break 3D CAPTCHAs without using OCR techniques by extracting pixel blocks from the characters of CAPTCHA challenges and using them for automated recognition \cite{nguyen2014security}.
Mori et al. utilized a shape context matching algorithm to crack Gimpy CAPTCHA with 33\% accuracy \cite{mori2003recognizing}.

Conti et al. summarized the main breaking strategies against image-based CAPTCHAs \cite{conti2016captchastar}. 
Some image-based CAPTCHAs just hide the solutions on the client-side, making them vulnerable to reverse engineering attacks. Others store pre-computed challenge answers in databases, thus prone to exhaustive searches or even random guesses if the number of responses is small. If such databases are compromised, these CAPTCHAs become easy prey.  
Golle employed a support vector machine classifier to automatically solve the Asirra CAPTCHA with a high success rate \cite{golle2008machine}. 
Instead of solving shape or image recognition, Hernandez-Castro et al. introduced a side-channel attack that leverages the JPEG file size to measure image continuity \cite{hernandez2015using}.

Sano et al. presented an attacking method against both audio and visual CAPTCHAs using Hidden Markov Models (HMMs) \cite{sano2015hmm}. The method could decode the audio and visual variants of Google’s reCAPTCHA with success rates of 58.75\% and 31.75\%, respectively. 
Kluever et al. proposed a video CAPTCHA mechanism and discussed a frequency-based attack that is effective against the proposed mechanism \cite{kluever2008video}.
The video CAPTCHA challenge requires users to submit three relevant tags as a solution. However, an automated program can exploit publicly available tag frequency estimation to break the mechanism.

\subsection{CAPTCHA Usability}
Yan et al. proposed a 3D framework to examine CAPTCHA usability that used distortion, content, and presentation to study various CAPTCHA types \cite{yan2008usability}. 
Sauer et al. studied the accessibility and usability of audio CAPTCHAs and found that their proposed scheme was accessible but had a low success rate of less than 90\% due to solvability challenges \cite{sauer2008towards}. 
Fidas et al. conducted a user survey to evaluate CAPTCHA usability \cite{fidas2011necessity}. The authors found that every other user needed at least two attempts to pass a CAPTCHA challenge, and that background patterns in CAPTCHAs had little impact on security and affected usability. 
Krol et al. compared and contrasted three CAPTCHA mechanisms: a traditional text-based reCAPTCHA, PlayThru, and NoBot \cite{krol2016better}. They focused on holistic user experience in solving CAPTCHAs. Time and workload were the two parameters used to compare the CAPTCHA mechanisms. They observed that using tablets resulted in significantly more hassles. 
Bursztein et al. led a study in which more than 300,000 CAPTCHAs were solved by workers on Amazon Mechanical Turk \cite{bursztein2010good}. The authors demonstrated that CAPTCHAs are hard for humans to solve, and audio-based CAPTCHAs are more tedious than their counterparts. 
Reynaga et al. presented users with nine CAPTCHA mechanisms with unconventional input schemes to enhance usability \cite{reynaga2015exploring}. The study, however, found that users preferred conventional methods due to familiarity. 
Gossweiler et al.'s study found dissatisfaction with text-based CAPTCHAs \cite{gossweiler2009s}, while Ho et al. experimented with CAPTCHA usability where participants had to solve CAPTCHAs several times \cite{ho2011deviltyper}. 

\subsection{Comparison with This Work}
Our work explores user perspectives on CAPTCHA security and usability, with a focus on the latter. We integrate usability issues identified in Yan et al.'s work \cite{yan2008usability} into our survey questions. 
Unlike prior studies focusing on specific CAPTCHA types (e.g., text-based \cite{yan2008usability,gossweiler2009s}, audio-based \cite{sauer2008towards}, or multiple types \cite{krol2016better,reynaga2015exploring}), we aim for a comprehensive overview encompassing all types.
The closest work to ours is Fidas et al.'s user survey \cite{fidas2011necessity}. Both studies address the user perceptions on security-usability trade-offs, attempts required to solve CAPTCHAs, language influences, etc. However, we want to investigate how user views have evolved since a decade ago as CAPTCHAs become more prevalent. Furthermore, we recruit participants from two distinct sources to see if their opinions vary significantly. Conversely, Fidas et al.'s study did not reveal where they recruited the 210 participants.

\section{Methodology}
\label{sec:method}
This study aimed to understand how users perceive and adopt CAPTCHA mechanisms.
To this end, we conducted an online survey to collect participants' views on the security and usability of CAPTCHAs.
It was reviewed and approved by our institution's Institutional Review Board.

\subsection{Recruitment}
We recruited participants from two sources: Amazon Mechanical Turk and university students majoring in Computer Science. 
We refer to these two groups as MTurk users and university users hereafter.

Amazon Mechanical Turk, also known as MTurk, is a marketplace where we can hire a diverse workforce to complete virtual tasks requiring human intelligence.
Compared to traditional participants in social science studies, MTurk users typically vary widely in age, income, education, and location \cite{henrich2010weirdest}. 
We selected candidates with a Human Intelligence Task (HIT) rate of over 90\% and more than 50 approved HITs for the study. 
Each participant was paid a \$1 incentive for completing the 5-minute survey.
Within 30 minutes, we received 150 responses, primarily from the United States.

University users were recruited from a public university in the United States. All participants in this group were enrolled in upper-division computer science courses and had a technical background.
 
\subsection{Study Design}
We used Qualtrics, a web-based survey tool, to conduct the anonymous survey. 
To avoid malpractice, participants would get a unique code at the end so they could only take the survey once.

Our questionnaire comprised 23 questions. 
The first seven questions gathered demographic information and determined survey eligibility. 
The next 16 questions focused on participants' perceptions of various CAPTCHA schemes regarding security and usability.

We first asked participants about different CAPTCHA types they had encountered daily, including math, text identification, image selection, image dragging or dropping, audio/video, and game CAPTCHAs. 
We also allowed participants to specify any other CAPTCHA types they experienced. 
We then asked participants to rate these different CAPTCHA types on a five-point Likert scale, from ``Least preferable'' (1) to ``Most preferable'' (5). 
Next, we asked participants if they had solved a CAPTCHA in a language other than English. If so, we asked them what type of non-English CAPTCHAs they preferred (e.g., text vs. audio/video). We allowed them to choose neither or both in this question.

The next few questions were about usability.
We first focused on small devices, such as a phone or a tablet. We asked participants if factors such as screen size, orientation, network bandwidth, and processor speed played a role in their perception. 
We then asked participants if, generally, they could solve a CAPTCHA in a single attempt. 
For those who generally took more than one attempt to solve a CAPTCHA, we followed up with an open-ended question to gather information about the reasons behind their difficulties.
We asked them if factors such as size, shape, color, distortion features, or background patterns made it harder for them to solve the challenge.  
We also designed some questions on users' frustration concerning solving CAPTCHA tests. 
We asked participants if they had ever abandoned a website because of a CAPTCHA challenge. 

As a security mechanism itself, the security of CAPTCHAs is also a significant concern. 
The last three questions attempted to understand users' perspectives on CAPTCHA security and overall effectiveness.

We screened potential participants based on their previous experience with CAPTCHAs and advanced only those who had solved a CAPTCHA at least once to the rest of the survey. We also recorded the time taken to complete the survey and discarded any results that were completed in less than 15 seconds. However, no participant was disqualified for this reason.
We received 154 responses from MTurk users, discarding four incomplete ones.
Meanwhile, all 109 responses from university users were valid.

\subsection{Demographics}
Before presenting our findings, we provide a summary of our respondents' demographics, outlined in Table \ref{table:demographics}.

\begin{table}[ht]
\caption{Demographics}
\footnotesize
\begin{center}
    \begin{minipage}[ht]{.44\linewidth}
        \begin{tabular}{p{2cm}llllll}
        \hline
                                              & \multicolumn{2}{c}{\textbf{MTurk}} & \multicolumn{2}{c}{\textbf{Univ.}} & \multicolumn{2}{c}{\textbf{Total}} \\
                                              & No.          & \%         & No.            & \%            & No.          & \%         \\ \hline
        \textbf{Gender}                                &              &            &                &               &              &            \\ \hline
        Male                                  & 82           & 55         & 90             & 83            & 172          & 66         \\
        Female                                & 68           & 45         & 15             & 14            & 83           & 32         \\
        Non-binary               & 0            & 0          & 2              & 2             & 2            & 1          \\
        Decline to say                     & 0            & 0          & 2              & 2             & 2            & 1          \\ \hline
        \textbf{Age}                                   &              &            &                &               &              &            \\ \hline
        18-24                                 & 6            & 4          & 65             & 60            & 71           & 27         \\
        25-34                                 & 69           & 46         & 38             & 35            & 107          & 41         \\
        35-44                                 & 38           & 25         & 5              & 5             & 43           & 17         \\
        45-54                                 & 19           & 13         & 1              & 1             & 20           & 8          \\
        55-64                                 & 12           & 8          & 0              & 0             & 12           & 5          \\
        64 or older                           & 6            & 4          & 0              & 0             & 6            & 2          \\ \hline
        \end{tabular}
    \end{minipage}
    \hfill
    \begin{minipage}[ht]{.5\linewidth}
        \begin{tabular}{p{3.4cm}llllll}
\hline
                                      & \multicolumn{2}{c}{\textbf{MTurk}} & \multicolumn{2}{c}{\textbf{Univ.}} & \multicolumn{2}{c}{\textbf{Total}} \\
                                      & No.          & \%         & No.            & \%            & No.          & \%         \\ \hline
        \textbf{Education}                             &              &            &                &               &              &            \\ \hline
        Less than high school degree          & 2            & 1          & 0              & 0             & 2            & 1          \\
        High school degree                    & 9            & 6          & 5              & 5             & 14           & 5          \\
        Some college experience               & 17           & 11         & 55             & 50            & 72           & 28         \\
        Associate degree                      & 15           & 10         & 36             & 33            & 51           & 20         \\
        Bachelor's degree                     & 85           & 57         & 13             & 12            & 98           & 38         \\
        Postgraduate degree                   & 22           & 15         & 0              & 0             & 22           & 8          \\ \hline
        \textbf{Language}                              &              &            &                &               &              &            \\ \hline
        Native English speaker                & 136          & 91         & 87             & 80            & 223          & 86         \\
        Non-native speaker            & 14           & 9          & 22             & 20            & 36           & 14         \\ \hline
        \end{tabular}
    \end{minipage}

\label{table:demographics}
\end{center}
\end{table}

Our survey included 259 valid respondents, with a higher proportion of males (62\%) compared to females (32\%). Among university users, only 14\% were female. Although we would have preferred a more balanced gender ratio in university users, these proportions align with the U.S. national average (18\%). 
The ages of our respondents ranged from 19 to 70 years old, with an average age of 32 and a standard deviation of 11.39. The largest age group was 25-34, constituting 41\% (107 out of 259) of respondents. Of these 107 respondents, 69 were MTurk users, and 38 were university users. 
Meanwhile, 60\% of university users (65 out of 109) were aged 18-24.
MTurk users had diverse educational backgrounds, with over 70\% holding a bachelor's degree or higher. 
In contrast, most university users (88\%) were pursuing their first bachelor's degree.
Among respondents, 86\% (223) were native English speakers. On the other hand, there are relatively more non-native English speakers in university users (20\%), reflecting the diversity of the university's student body.

\subsection{Hypotheses}
The following null hypotheses were formulated in this study:
\begin{itemize}
    \item (H1) There is no significant difference in the user perception of preferred CAPTCHA types between the two user groups.
    \item (H2) There is no significant difference in the user perception of solving CAPTCHAs on smaller devices between the two user groups.
    \item (H3) Both user groups abandon CAPTCHA challenges for similar reasons.
    \item (H4) There is no significant difference in the user perception of CAPTCHA difficulty between the two user groups.
    \item (H5) There is no significant difference in the user perception of CAPTCHA security between the two user groups. 
    \item (H6) There is no significant difference in overall opinion of CAPTCHAs between the two user groups. 
\end{itemize}

\section{Findings}
\label{sec:findings}
After gathering the responses, we analyzed the data from the Likert and closed-ended questions.  
We wanted to determine if there is any difference between the two distinct user groups, MTurk and university users, regarding their perspectives on CAPTCHAs.

\subsubsection{Use and preferences on CAPTCHA types}
\begin{figure}[htb]
\begin{center}
\begin{tikzpicture}[scale = 0.75]
  \begin{axis}[
    ybar stacked, ymin=0,  
    bar width=7mm,
    symbolic x coords={Math, Text identification, Image recognition, Image drag and drop, Audio or video, Game or puzzle},
    xtick=data,
    nodes near coords, 
    nodes near coords align={anchor=north},
    totals/.style={nodes near coords align={anchor=south}},
    x tick label style={anchor=south,font=\scriptsize,yshift=-1cm,text width=1cm,align=center},
  ]
  \addplot [fill=red!50] coordinates {
({Math},76)
({Text identification},97)
({Image recognition},124)
({Image drag and drop},46)
({Audio or video},39) 
({Game or puzzle},37)};
  \addplot [fill=yellow!50,point meta=explicit] coordinates {
({Math},35) [35]
({Text identification},92) [92]
({Image recognition},100) [100]
({Image drag and drop},29) [29]
({Audio or video},18) [18]
({Game or puzzle},31) [31]};
  \addplot[totals] coordinates {
({Math},0)
({Text identification},0)
({Image recognition},0)
({Image drag and drop},0)
({Audio or video},0) 
({Game or puzzle},0)};
  \legend{\strut MTurk, \strut University}
  \end{axis}
\end{tikzpicture}
\end{center}
\caption{Use of different CAPTCHAs (n=253)}
\label{fig:use-of-captchas}
\end{figure}
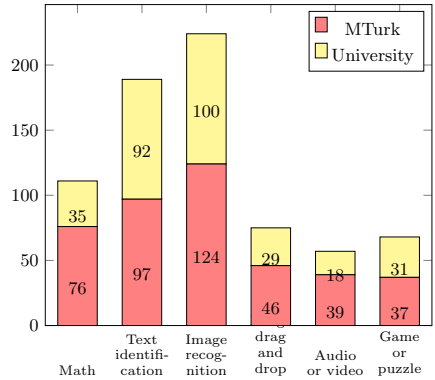

We surveyed respondents on their use of different CAPTCHA types, including math, text identification, image recognition, image drag and drop, audio/video, and game/puzzle.
Figure \ref{fig:use-of-captchas} shows that image recognition is the most prevalent type (224 votes, 89\%) for both user groups, followed by text identification CAPTCHAs (189 votes, 75\%). The least used were audio/video CAPTCHAs (57 votes, 23\%). 
These results align with current market trends, with many websites still using Google reCAPTCHA v2 or earlier versions, featuring image recognition challenges. 
Text identification CAPTCHAs remain competitive, as websites like Amazon and Facebook continue to utilize them. 
However, audio/video CAPTCHAs, initially designed for users with disabilities, face practical challenges due to accessibility and availability issues.

We then asked respondents to rank their preferences for different CAPTCHAs on a scale of 1 to 5, with 1 being least preferable and 5 being the most preferable. 
Figure \ref{fig:mturk-preferences} shows the preferences of MTurk users, and Figure \ref{fig:univ-preferences} presents those of university users. 
Text identification is the most preferred CAPTCHA among MTurk users (55 votes, 37\%), followed by image recognition (45 votes, 31\%). 
Conversely, university users favored image recognition the most (33 votes, 31\%), with text identification as their second choice (23 votes, 22\%).

\begin{figure}[htbp]
\centering

\begin{minipage}{0.5\textwidth}

\pgfplotstableread[col sep=comma,header=true]{
Captcha,1,2,3,4,5,6
Math,13,13,32,21,38,30
Text identification,8,18,30,27,55,9
Image recognition,27,16,17,26,45,16
Image drag and drop,12,14,17,28,31,45
Audio or video,19,15,18,22,23,50
Game or puzzle,17,12,12,26,36,44
}\data

\pgfplotstablecreatecol[
 create col/expr={
    \thisrow{1} + \thisrow{2} + \thisrow{3} +\thisrow{4} + \thisrow{5} + \thisrow{6}
 }
]{sum}{\data}

\pgfplotsset{
  percentage plot/.style={
    point meta=explicit,
    every node near coord/.append style={
      font=\scriptsize,
      color=black,
    },
    nodes near coords={
      \pgfmathtruncatemacro\iszero{\originalvalue==0}
      \ifnum\iszero=0
      \pgfmathprintnumber[fixed,fixed zerofill,precision=0]{\pgfplotspointmeta}
      \fi
    },
    yticklabel=\pgfmathprintnumber{\tick}\,$\%$,
    ymin=0,
    ymax=100.01, 
    visualization depends on={y \as \originalvalue},
    enlarge x limits={abs=6mm}
  },
  percentage series/.style={
    table/x expr=\coordindex, 
    table/y expr=(\thisrow{#1}/\thisrow{sum}*100),
    table/meta=#1
    }
}

\begin{tikzpicture}[scale = 0.75]
\begin{axis}[
    ybar stacked,
    percentage plot,
    bar width=0.55cm, 
    xticklabels from table={\data}{Captcha}, 
    xtick=data,
    x tick label style={
      rotate=45,
      anchor=east,
      font=\scriptsize,
      xshift=-1.5mm, yshift=-2mm
    },
    legend style={
      at={(0.5,-0.4)},
      anchor=south,
      legend columns=-1
      },
]

    \addplot [fill=red!50]   table[percentage series=1] {\data};
    \addplot [fill=orange!50]  table[percentage series=2] {\data};
    \addplot [fill=yellow!50]  table[percentage series=3] {\data};
    \addplot [fill=green]       table[percentage series=4] {\data};
    \addplot [fill=blue!30]       table[percentage series=5] {\data};
    \addplot [fill=violet!30]       table[percentage series=6] {\data};


    \legend{1, 2, 3, 4, 5, Never used}
\end{axis}
\end{tikzpicture}
\caption{Preferences of MTurk users}
\label{fig:mturk-preferences}
\end{minipage}\hfill
\begin{minipage}{0.5\textwidth}


\pgfplotstableread[col sep=comma,header=true]{
Captcha,1,2,3,4,5,6
Math,14,10,14,6,11,51
Text identification,17,14,22,26,23,4
Image recognition,18,15,21,19,33,0
Image drag and drop,14,10,9,10,8,55
Audio or video,21,8,6,5,1,65
Game or puzzle,12,7,11,10,10,56
}\data

\pgfplotstablecreatecol[
 create col/expr={
    \thisrow{1} + \thisrow{2} + \thisrow{3} +\thisrow{4} + \thisrow{5} + \thisrow{6}
 }
]{sum}{\data}

\pgfplotsset{
  percentage plot/.style={
    point meta=explicit,
    every node near coord/.append style={
      font=\scriptsize,
      color=black,
    },
    nodes near coords={
      \pgfmathtruncatemacro\iszero{\originalvalue==0}
      \ifnum\iszero=0
      \pgfmathprintnumber[fixed,fixed zerofill,precision=0]{\pgfplotspointmeta}
      \fi
    },
    yticklabel=\pgfmathprintnumber{\tick}\,$\%$,
    ymin=0,
    ymax=100.01, 
    visualization depends on={y \as \originalvalue},
    enlarge x limits={abs=6mm}
  },
  percentage series/.style={
    table/x expr=\coordindex, 
    table/y expr=(\thisrow{#1}/\thisrow{sum}*100),
    table/meta=#1
    }
}

\begin{tikzpicture}[scale = 0.75]
\begin{axis}[
    ybar stacked,
    percentage plot,
    bar width=0.55cm, 
    xticklabels from table={\data}{Captcha}, 
    xtick=data,
    x tick label style={
      rotate=45,
      anchor=east,
      font=\scriptsize,
      xshift=-1.5mm, yshift=-2mm
    },
    legend style={
      at={(0.5,-0.4)},
      anchor=south,
      legend columns=-1
      },
]

    \addplot [fill=red!50]   table[percentage series=1] {\data};
    \addplot [fill=orange!50]  table[percentage series=2] {\data};
    \addplot [fill=yellow!50]  table[percentage series=3] {\data};
    \addplot [fill=green]       table[percentage series=4] {\data};
    \addplot [fill=blue!30]       table[percentage series=5] {\data};
    \addplot [fill=violet!30]       table[percentage series=6] {\data};


    \legend{1, 2, 3, 4, 5, Never used}
\end{axis}
\end{tikzpicture}
\caption{Preferences of university users}
\label{fig:univ-preferences}
\end{minipage}
\end{figure}

We also observed that over half of university users never solved any of the other four CAPTCHA types, whereas MTurk users showed a more diverse usage pattern. 
Additionally, a significant difference was observed between the two groups in their CAPTCHA preferences (p-value $<$ 0.05), as determined by an Analysis of Variance (ANOVA) test. 
This disparity could be due to several factors, including age, as most university users were under 25 years old. 

\subsubsection{CAPTCHAs in non-English languages} 
In our survey, only 3\% of respondents reported encountering a non-English CAPTCHA, with Portuguese being the most common language at 37.5\%, followed by other languages like Hindi, Tamil, Turkish, French, Japanese, and Korean.
Among these respondents, 50\% preferred text-based CAPTCHAs, while 13\% favored audio/video CAPTCHAs, and 25\% preferred both. However, due to the limited sample size, further research is needed to understand user preferences in this area.

\subsubsection{Solving CAPTCHAs on a smaller device}
To investigate the impact of device size on CAPTCHA-solving, we asked respondents about their preferences. 
Our results show that 68\% of respondents found it easier on a computer than on a phone or a tablet, while 32\% perceived no difference. Notably, our analysis revealed a significant difference in opinions between MTurk and university users, with a chi-squared p-value of 0.0004. This finding indicates the need for additional research to explore and address usability issues on small devices.

\begin{figure}
    \begin{minipage}[htb]{.4\linewidth}
        \captionof{table}{Solving CAPTCHAs on a small device}
        \footnotesize
        \centering
        \begin{tabular}{p{1.75cm}lll}
                                 & MTurk      & Univ. & Total \\ \hline
        Easier on a computer     & 114 (78\%) & 58 (55\%)  & 172   \\ \hline
        Easier on a phone/tablet & 0 (0\%)    & 1 (1\%)    & 1     \\ \hline
        Same                     & 33 (22\%)  & 47 (44\%)  & 253   \\ \hline
        \end{tabular}
        \label{table:small-devices}   
    \end{minipage}
    \hfill
    \begin{minipage}[htb]{.45\linewidth}
        \centering
        \begin{tikzpicture}[scale = 0.7]
        \pgfplotstableread{
            X   MTurk   University   
            {Screen size and orientation}   82   46   
            {Input}   59    42
            {Network bandwidth} 25  9
            {Processor speed}   23  10 
            {Other} 1   1
        }\data
        
        \begin{axis}[
            ybar=1mm,     
            bar width=3mm,  
            ymin=0, ymax=100,
            xtick=data,
            xticklabels from table = {\data}{X},
            x tick label style={anchor=south,font=\scriptsize,yshift=-1cm,text width=1.5cm,align=center},
            legend style = {legend columns=-1,
                            legend pos=north west,
                            font=\footnotesize,
                            /tikz/every even column/.append style={column sep=2mm},
                            },
            ]
        \addplot table[x expr=\coordindex,y index=1] {\data};
        \addplot table[x expr=\coordindex,y index=2] {\data};
        
        \legend{MTurk, University}
        \end{axis}
        \end{tikzpicture}
        
        \caption{Reasons for choosing computers over small devices (n=172)}
        \label{fig:small-devices}
    \end{minipage}

\end{figure}

We received feedback from 172 respondents on why they found using one device easier than the other. The majority (74\%) cited size and orientation as the main reasons, with 73\% from MTurk users and 78\% from university users. Additionally, over half (58\%) considered input method (Touch vs. Keyboard/Mouse) as an essential factor. This information is depicted in Figure \ref{fig:small-devices}.

\subsubsection{Failing a CAPTCHA challenge on a single attempt}
\label{sec:failing-one-attempt}
Solving CAPTCHAs can be a frustrating experience for users.
In our survey, we asked respondents about their success rate in solving CAPTCHAs on the first attempt. We found that a staggering 28\% of respondents could always solve a CAPTCHA challenge on the first try, as shown in Table \ref{table:one-attempt-success}. 
However, 3\% of respondents consistently struggled with CAPTCHAs. We also asked them to share their experiences of failure in an open-ended question.

\begin{figure}
    \begin{minipage}[htb]{.4\linewidth}
        \captionof{table}{Solving CAPTCHAs in one attempt}
        \footnotesize
        \centering
        \begin{tabular}{lllll}
                                                         &    & MTurk & Univ. & Total \\ \hline
        \multicolumn{1}{l|}{\multirow{2}{*}{Always}}     & n  & 52    & 19         & 71    \\ \cline{2-5} 
        \multicolumn{1}{l|}{}                            & \% & 35\%  & 18\%       & 28\%  \\ \hline
        \multicolumn{1}{l|}{\multirow{2}{*}{Frequently}} & n  & 90    & 85         & 175   \\ \cline{2-5} 
        \multicolumn{1}{l|}{}                            & \% & 61\%  & 80\%       & 69\%  \\ \hline
        \multicolumn{1}{l|}{\multirow{2}{*}{Rarely}}     & n  & 5     & 2          & 7     \\ \cline{2-5} 
        \multicolumn{1}{l|}{}                            & \% & 3\%   & 2\%        & 3\%   \\ \hline
        \multicolumn{1}{l|}{}                            &    & 147   & 106        & 253   \\ \hline
        \end{tabular}
        \label{table:one-attempt-success}
    \end{minipage}
    \hfill
    \begin{minipage}[htb]{.45\linewidth}
        \centering
        \begin{tikzpicture}[scale = 0.7]
        \pgfplotstableread{
            X   MTurk   University   
            {Distortion}   98   75   
            {Size, shape, and color}   68    40
            {Background patterns} 58  37
            {Other} 6   11
        }\data
        
        \begin{axis}[
            ybar=1mm,     
            bar width=3mm,  
            ymin=0, ymax=100,
            xtick=data,
            xticklabels from table = {\data}{X},
            x tick label style={anchor=south,font=\scriptsize,yshift=-1cm,text width=1.5cm,align=center},
            legend style = {legend columns=-1,
                            legend pos=north west,
                            font=\footnotesize,
                            /tikz/every even column/.append style={column sep=2mm},
                            },
            ]
        \addplot table[x expr=\coordindex,y index=1] {\data};
        \addplot table[x expr=\coordindex,y index=2] {\data};
        
        \legend{MTurk, University}
        \end{axis}
        \end{tikzpicture}
        
        \caption{Why CAPTCHAs are hard to use?}
        \label{fig:hard-to-use}
    \end{minipage}

\end{figure}

Respondents commonly cited image blurriness or unclear visuals as a primary issue. 
For instance, respondent P2 complained: \textit{``I can't always see the images clearly.''} 
P122 stated: \textit{``The picture was too dark fuzzy or unclear.''} 
P17 commented: \textit{``Pictures too blurred or small to see properly.''}

In addition, text clarity was also cited as a problem, with respondents stating that the font, font size, and text formatting could have been easier to read.
For example, P25 said: \textit{``Cannot recognize font.''} 
P197 shared: \textit{``CAPTCHA text or image can be distorted beyond recognition.''}
P101 echoed this opinion: \textit{``A lot of times the text or number identification for Amazon logins is really unclear.''} 
P14 also wrote: \textit{``The text ones are sometimes really difficult to read.''}
Some respondents reported having to refresh the CAPTCHA test for a clearer one.

Furthermore, the perceived similarity of letters and numbers in CAPTCHAs was another highly cited reason for failure. 
For example, respondents had trouble distinguishing letters like `9' vs. `g,' ``vv'' vs. `w,' `I' vs. `1.'
P73 explained: \textit{``Some letter number combos like nine g and one I l and s f and o zero and similar letter number combo are very difficult to solve.''}
P53 shared the same concern: \textit{``With text identification, it is sometimes difficult to ascertain the letter correctly like a letter l and number one can be similar.''}
These similarities often led to typing errors and failures in solving CAPTCHAs on the first attempt.

Human errors are inevitable and have caused many user failures when interacting with CAPTCHAs. 
Some users did not spend sufficient time on solving CAPTCHAs, while others gave inadequate attention to CAPTCHA tests. 
These errors are particularly prevalent in image recognition and text identification CAPTCHAs, which prevail in the market.
P259 shared their experience, stating: \textit{``Mistaken letter or misheard a letter.''}
Similarly, P32 mentioned difficulties encountered in text CAPTCHAs, noting: \textit{``Sometimes miss a picture or put a letter in the wrong place in a text CAPTCHA.''}
These challenges are not unique to them. P89 struggled with image CAPTCHAs, expressing frustration: \textit{``Usually, it's on the image ones and I just miss one of the items or squares it wants me to click.''}
Users often face difficulty deciding whether to select a particular image segment when only a small portion of the target object is visible. 
P163 highlighted a common error: \textit{``Normally incorrect image selected or character.''}
P33 also expressed their frustrations, stating: \textit{``I did not select all the pictures.''}
Responses such as ``carelessness,'' ``typo,'' ``accidental click,'' and ``rushed'' were also common. 

Some text CAPTCHAs are case-sensitive, which posed a challenge for participants who struggled to determine the required accuracy. As a result, they often had to attempt the test multiple times before passing. 
P64 talked about such experience, stating: \textit{``Typing a lowercase while CAPTCHA needs an uppercase.''}
This issue commonly leads to another problem: unclear instructions. 
Many respondents observed vague or incomplete instructions in the CAPTCHAs they tried to solve. 
P181 recounted their experience, saying: \textit{``The image says to click all the stoplights, and I don't know if it wants me to click the little bit on a square that is part of it but not really the stoplights.''}
Additionally, P129 expressed frustration with the system rejecting seemingly correct answers, stating: \textit{``I don't know. I type the exact text from the image, and then it says it's wrong. The first one is probably always expired or something.''}
Our findings suggest that intentionally obscuring instructions may be a deliberate choice to enhance CAPTCHA's difficulty for bots.

\subsubsection{Why did you give up?}
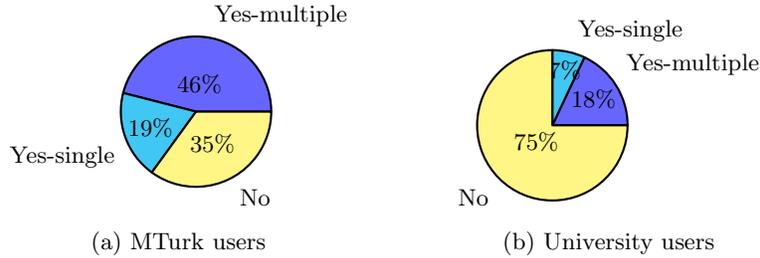
\begin{figure}[htbp]
\centering
\begin{subfigure}[t]{0.46\linewidth}
    \centering
    \begin{tikzpicture}
        \pie[radius=1]{46/{Yes-multiple}, 19/{Yes-single}, 35/No}
    \end{tikzpicture}
    \caption{MTurk users}
\end{subfigure}
\begin{subfigure}[t]{0.46\linewidth}
    \centering
    \begin{tikzpicture}
        \pie[radius=1]{18/{Yes-multiple}, 7/{Yes-single}, 75/No}
    \end{tikzpicture}
    \caption{University users}
\end{subfigure}
\caption{Leaving a webpage after failed attempts}
\label{fig:failed-attempts}
\end{figure}

We investigated if respondents ever left a webpage due to CAPTCHA failures. 
Figure \ref{fig:failed-attempts} illustrates their tendency to abandon after multiple failed attempts or just a single attempt. 
Surprisingly, 75\% of university users would persist until success. 
In contrast, nearly 65\% of MTurk users opted to abandon, with 19\% leaving immediately after a single failure compared to 7\% of university users.

To gain further insights, we posed a related question, aiming to ascertain if users left a webpage because they simply did not want to solve a CAPTCHA.  
Among university users, 62\% reported that they would remain on the webpage despite the presence of a CAPTCHA.  
In contrast, nearly 65\% of MTurk users admitted to leaving a webpage due to finding CAPTCHAs challenging or annoying, or discovering alternative bypass methods.

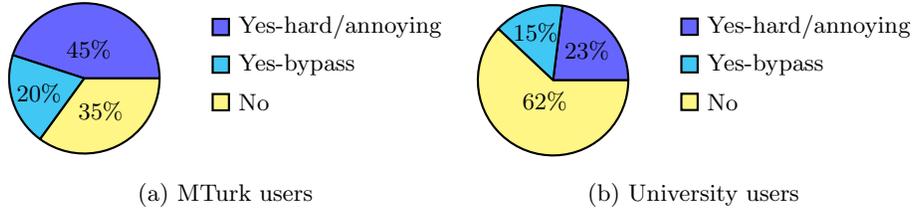
\begin{figure}[htbp]
\centering
\begin{subfigure}[t]{0.49\textwidth}
    \centering
    \begin{tikzpicture}
        \pie[radius=1,text=legend]{45/{Yes-hard/annoying}, 20/{Yes-bypass}, 35/No}
    \end{tikzpicture}
    \caption{MTurk users}
\end{subfigure}
\hfill
\begin{subfigure}[t]{0.49\textwidth}
    \centering
    \begin{tikzpicture}
        \pie[radius=1,text=legend]{23/{Yes-hard/annoying}, 15/{Yes-bypass}, 62/No}
    \end{tikzpicture}
    \caption{University users}
\end{subfigure}
\caption{Leaving a webpage when a CAPTCHA is present}
\label{fig:left-captcha-pages}
\end{figure}

The two user groups exhibited significantly different behaviors in response to these questions (p $<$ 0.0001, chi-squared test). We believe that active MTurk users tend to complete tasks or jobs as much as possible within a given time frame. They probably extend this mindset to their browsing experiences and have little patience towards CAPTCHAs. 
Conversely, university students, being tech-savvy and goal-oriented, show more patience during the authentication process. 

\subsubsection{Perceived CAPTCHA difficulty}
The next two questions address the perceived difficulty of CAPTCHAs. 
We asked respondents to select the feature that contributes most to CAPTCHA difficulty. Figure \ref{fig:hard-to-use} presents the distribution of votes from our participants. 
Distortion emerged as the primary factor that makes CAPTCHAs challenging to use, aligning with the insights gathered from user responses in Section \ref{sec:failing-one-attempt}. Notably, 98 MTurk users (65\%) and 75 university users (69\%) endorsed this option. 
Size, shape, and color were also identified as troublesome factors, with 68 votes from MTurk users (45\%) and 40 votes from university users (37\%). Background patterns ranked third, albeit with a narrow margin (58 votes from MTurk users and 37 votes from university users).

\begin{figure}[htbp]
\centering
\begin{subfigure}[t]{0.46\linewidth}
    \centering
    \begin{tikzpicture}
        \pie[radius=1]{27/{Easier}, 43/{Harder}, 30/{The same}}
    \end{tikzpicture}
    \caption{MTurk users}
\end{subfigure}
\begin{subfigure}[t]{0.46\linewidth}
    \centering
    \begin{tikzpicture}
        \pie[radius=1]{19/{Easier}, 32/{Harder}, 49/{The same}}
    \end{tikzpicture}
    \caption{University users}
\end{subfigure}
\caption{Trend of CAPTCHA difficulty}
\label{fig:trend}
\end{figure}
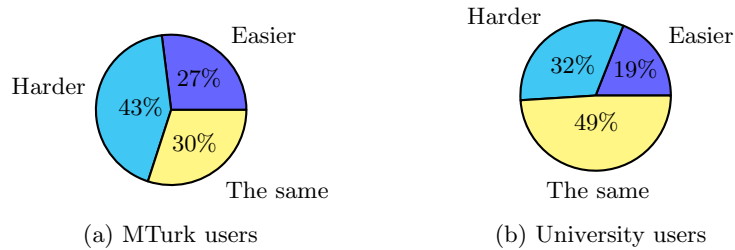

Next, we inquired about respondents' perceptions of the trend in CAPTCHA difficulty over time. As illustrated in Figure \ref{fig:trend}, 43\% of MTurk users felt that CAPTCHAs are becoming increasingly hard to solve, while 49\% of university users believed that CAPTCHA difficulty has remained consistent. A chi-squared test of independence revealed significant differences between the responses of the two user groups (p = 0.008). 

\subsubsection{Accessibility concerns}
Accessibility is crucial for user experience, yet several respondents shared frustrations with CAPTCHA accessibility. 
P12 noted that \textit{``audio files sometimes did not properly load when solving an audio CAPTCHA.''} This issue may be attributed to poor Internet connectivity.
P262 reported an unpleasant experience with audio CAPTCHAs: \textit{``Helping a customer in which they had a hard time in hearing.''} However, without further information, we were not sure about the reason behind this incident.
Additionally, P181 encountered issues when using small mobile devices, particularly in touchscreen mode. They stated, \textit{``I was on a mobile phone and tried to solve a CAPTCHA and when I couldn't click on the screen, I chose to make it an audio CAPTCHA and it failed to let me solve it.''}
This complaint echoes the general sentiment among our respondents regarding CAPTCHAs on small screens.

\subsubsection{How secure are current CAPTCHAs?}
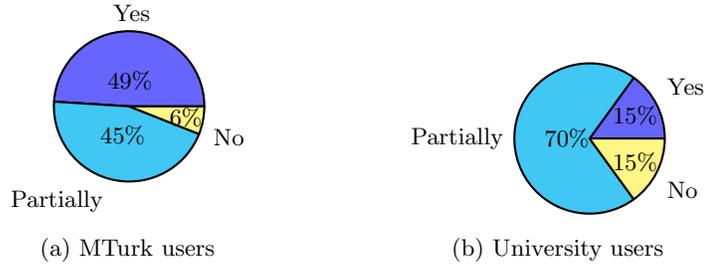
\begin{figure}[ht]
\centering
\begin{subfigure}[t]{0.46\linewidth}
    \centering
    \begin{tikzpicture}
        \pie[radius=1]{49/{Yes}, 45/{Partially}, 6/{No}}
    \end{tikzpicture}
    \caption{MTurk users}
\end{subfigure}
\begin{subfigure}[t]{0.46\linewidth}
    \centering
    \begin{tikzpicture}
        \pie[radius=1]{15/{Yes}, 70/{Partially}, 15/{No}}
    \end{tikzpicture}
    \caption{University users}
\end{subfigure}
\caption{Can CAPTCHAs detect bots?}
\label{fig:bot-elimination}
\end{figure}

We also assessed users' perceptions of CAPTCHA security.  
The results, shown in Figure \ref{fig:bot-elimination}, indicate that 49\% of MTurk users praised the effectiveness of CAPTCHAs, while 45\% agreed that CAPTCHAs do a decent job. Conversely, only 15\% of university users expressed high confidence in CAPTCHA effectiveness, with 70\% thinking CAPTCHAs do an okay job.  

\begin{figure}[htbp]
\centering
\begin{subfigure}[t]{0.46\linewidth}
    \centering
    \begin{tikzpicture}
        \pie[radius=1]{44/{Yes}, 15/{No}, 41/{Not sure}}
    \end{tikzpicture}
    \caption{MTurk users}
\end{subfigure}
\begin{subfigure}[t]{0.46\linewidth}
    \centering
    \begin{tikzpicture}
        \pie[radius=1]{25/{Yes}, 7/{No}, 69/{Not sure}}
    \end{tikzpicture}
    \caption{University users}
\end{subfigure}
\caption{Are CAPTCHAs resistant to attacks?}
\label{fig:attack-resistance}
\end{figure}
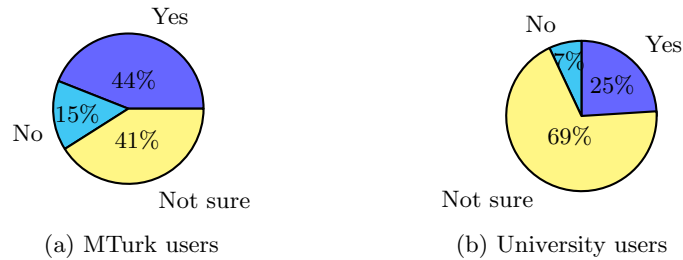

Next, we inquired about respondents' views on whether current CAPTCHA schemes can keep up with the advancing attack techniques.
Figure \ref{fig:attack-resistance} reveals that 69\% of university users were uncertain. 
Similarly, 41\% of MTurk users shared this uncertainty, while 44\% expressed confidence in CAPTCHA security.  
These findings suggest that university users are more aware of security threats around us since many of them have taken security courses and are tech-savvy. 
However, they also highlight the need for greater public awareness and education on cybersecurity and digital threats.

\subsubsection{Final stance}
The final question aimed to gather respondents' overall opinions regarding the security and usability of current CAPTCHAs. 
We found that 44\% of university users deemed CAPTCHAs secure, while 66\% found them usable. 
In contrast, MTurk users exhibited higher levels of confidence in CAPTCHA security, with 64\% considering them secure and approximately 55\% perceiving them as usable. 
Only a small proportion of MTurk users (6\%) and university users (8\%) viewed CAPTCHAs as neither secure nor usable.  
Once again, a significant disparity emerged between the overall opinions of the two user groups regarding CAPTCHAs (p = 0.021, chi-squared test).

\begin{table}[t]
\caption{Final stance on CAPTCHAs}
\footnotesize
\begin{center}
\begin{tabular}{lll}
                                  & MTurk  & University   \\ \hline
Secure and usable                 & 35 (24\%)  & 20 (19\%)        \\ \hline
Secure but usability is a concern & 57 (40\%)  & 27 (25\%)        \\ \hline
Usable but security is a concern  & 43 (30\%)  & 50 (47\%)        \\ \hline
Neither secure nor usable         & 9 (6\%)    & 9 (8\%)          \\ \hline
\end{tabular}
\label{table:final-stance}
\end{center}   
\end{table}

\section{Discussions}
\label{sec:discussions}
All six null hypotheses in our study were rejected, indicating significant variations in user perceptions of CAPTCHA issues between MTurk users and university users.

We observed distinct preferences for CAPTCHA types and differing perspectives on CAPTCHA difficulty trends among the two user groups.
While more MTurk users claimed to always solve CAPTCHAs on the first attempt (35\% vs. 18\%), they also tended to give up on CAPTCHA challenges compared to university users, who demonstrated greater persistence and patience.
User age played a role in these differences. Unlike Fidas et al.'s study \cite{fidas2011necessity}, we did not categorize respondents by specific age groups. 
However, our findings indicated that MTurk users, who were generally older (96\% aged 25 or older), differed from university users, of whom the majority (60\%) were under 25 years old. 
Our results also aligned with previous research \cite{reynaga2015exploring} that highlighted preferences for different CAPTCHA schemes on smartphones among younger individuals. 
Moreover, our study corroborated the finding that older participants are more likely to solve CAPTCHAs on the first attempt \cite{fidas2011necessity}.  

Contrary to common belief, MTurk users in our study, despite having a higher level of education (72\% holding a bachelor's or postgraduate degree), exhibited less concern for CAPTCHA security (64\% considered it secure). 
We attribute this discrepancy to the influence of technical background rather than education level. Our university users, mainly from a computer security class, displayed greater awareness of increasing security threats on the Internet. 
 
Both MTurk users and university users found solving CAPTCHAs on smaller devices equally challenging (100\% for MTurk users and 99\% for university users). 
Screen size, orientation, and input methods were identified as the top obstacles in this context. 
Many general CAPTCHA issues persisted and may even be exacerbated on smaller devices, including distortion, font size, and background patterns. 
To address these concerns, we support the recommendations by Reynaga et al. \cite{reynaga2015exploring} that CAPTCHAs on smaller devices should adopt input mechanisms with cross-platform support, maintain a consistent orientation, and employ a minimalist design to reduce screen occupancy.
 
Our study reveals that CAPTCHAs remain difficult to use. 
Moreover, current CAPTCHA designs heavily rely on visual processing, rendering them inaccessible to millions of people with visual impairments \cite{fanelle2020blind}. 
There is a growing demand for simpler and less burdensome solutions.
Google's reCAPTCHA v2 introduced \textit{Invisible CAPTCHAs} in 2014, but privacy concerns persist. 
We believe an invisible and passive CAPTCHA scheme with a frictionless user experience is the right direction.
Such a design frees users from the need to solve CAPTCHA challenges directly. It leverages the nature of a user's device and behavioral activities (e.g., CPU usage, memory usage, keystrokes, mouse movement, browser tabs, etc.). It should be accessible to everyone and less annoying. More importantly, it needs to be transparent about the use of user information. 
Our future agenda involves developing and testing such a new mechanism to address identified concerns.

\section{Conclusion}
\label{sec:conclusion}
In this study, we surveyed 259 respondents from a university and Amazon Mechanical Turk to examine user perceptions of CAPTCHAs, focusing on usability and security.
Our findings revealed significant disparities between the two groups. We explored possible explanations for these differences, including age and technical background.
Both groups struggled with current CAPTCHA schemes, leading to failure and frustration. 
Our insights offer recommendations for designing more user-friendly CAPTCHAs. 
Future research can use our findings to develop more effective and efficient solutions.

\begin{subappendices}
\renewcommand{\thesection}{\Alph{section}}%
\section{Survey Questions}\label{appendix:survey}

\Qitem{ \Qq{What is your age?}}

\Qitem{ \Qq{What is your gender?}
\begin{Qlist}
    \item Male
    \item Female
    \item Non-binary
    \item Prefer not to say
\end{Qlist}
}

\Qitem{ \Qq{What is the highest level of education that you have completed?}
\begin{Qlist}
    \item Less than high school degree
    \item High school degree or equivalent (e.g., GED)
    \item Some college but no degree
    \item Associate degree
    \item Bachelor's degree
    \item Graduate degree
\end{Qlist}
}

\Qitem{ \Qq{Which of the following best describes your current occupation?}
\begin{Qlist}
    \item Unemployed
    \item Self-employed
    \item Student
    \item Education
    \item Healthcare
    \item Information technology
    \item Real estate and development
    \item Retail
    \item Law
    \item Agriculture
    \item Government
    \item Other
\end{Qlist}
}

\Qitem{ \Qq{Which of the following best describes you?}
\begin{Qlist}
    \item Native English speaker
    \item Non-Native English speaker
\end{Qlist}
}

\Qitem{ \Qq{When was the last time you solved a CAPTCHA? [Skip to end survey if ``I have never solved a CAPTCHA'' is selected]}
\begin{Qlist}
    \item Today
    \item In the last week
    \item In the last two weeks
    \item In the last month
    \item In the last six months
    \item In the last year
    \item It has been over a year since I solved a CAPTCHA
    \item I have never solved a CAPTCHA
\end{Qlist}
}

\Qitem{ \Qq{Have you solved the below CAPTCHA types? Select all that applies.}
\begin{Qlist}
    \item Math CAPTCHA
    \item Text Identiciation CAPTCHA
    \item Select Images CAPTCHA
    \item Drag or Drop Images CAPTCHA
    \item Audio/Video CAPTCHA
    \item Game/Puzzle CAPTCHA
    \item Other (Please Specify) \Qline{1.5cm}
\end{Qlist}
}

\Qitem{ \Qq{Rate these CAPTCHA types on a scale of 1 - 5 where 1 is least preferable to you and 5 being the most preferable to you.}
\begin{Qlist}
    \item Math CAPTCHA (1, 2, 3, 4, 5, Never solved)
    \item Text Identiciation CAPTCHA (1, 2, 3, 4, 5, Never solved)
    \item Select Images CAPTCHA (1, 2, 3, 4, 5, Never solved)
    \item Drag or Drop Images CAPTCHA (1, 2, 3, 4, 5, Never solved)
    \item Audio/Video CAPTCHA (1, 2, 3, 4, 5, Never solved)
    \item Game/Puzzle CAPTCHA (1, 2, 3, 4, 5, Never solved)
\end{Qlist}
}

\Qitem{ \Qq{Have you solved a CAPTCHA in a language other than English? If yes, please specify all the languages.}
\begin{Qlist}
    \item Languages \Qline{3cm}
    \item No
\end{Qlist}
}

\Qitem{ \Qq{If you ever solved a non-English CAPTCHA, what type do you prefer?}
\begin{Qlist}
    \item Audio/Video
    \item Text
    \item Neither
    \item Both
\end{Qlist}
}

\Qitem{ \Qq{Do you think solving a CAPTCHA is easier on a computer than on a phone/tablet?}
\begin{Qlist}
    \item Easier on a computer.
    \item Easier on a phone/tablet.
    \item Same in both cases.
\end{Qlist}
}

\Qitem{ \Qq{In the last question, what is the reason for selecting one over the other? Select all that applies.}
\begin{Qlist}
    \item Screen size and orientation
    \item Input (touch vs. keyboard/mouse)
    \item Network bandwidth
    \item Processor speed
    \item Other (Please explain) \Qline{1.5cm}
\end{Qlist}
}

\Qitem{ \Qq{Generally, are you able to solve a CAPTCHA in a single attempt?}
\begin{Qlist}
    \item Always
    \item Frequently
    \item Rarely
    \item Never
\end{Qlist}
}

\Qitem{ \Qq{Please explain the reason for failing in the first attempt. \Qline{1.5cm}}}

\Qitem{ \Qq{What according to you makes a CAPTCHA hard to use?}
\begin{Qlist}
    \item Distortion feature
    \item Size, shape, and color
    \item Background patterns
    \item Other (Please explain) \Qline{1.5cm}
\end{Qlist}
}

\Qitem{ \Qq{What are your thoughts about the CAPTCHA trend over the years?}
\begin{Qlist}
    \item They are getting harder to solve.
    \item They are getting easier to solve.
    \item They are the same.
\end{Qlist}
}

\Qitem{ \Qq{Did you ever leave a webpage because you could not solve a CAPTCHA? Select all that applies.}
\begin{Qlist}
    \item Yes, I had to leave because I could not solve after multiple attempts.
    \item Yes, I left because I could not solve in a single attempt.
    \item No
\end{Qlist}
}

\Qitem{ \Qq{Did you ever leave a webpage because you did not want to solve a CAPTCHA? Select all that applies.}
\begin{Qlist}
    \item Yes, I left because the CAPTCHA looked hard/annoying to solve.
    \item Yes, I left hoping I could find the same information on a different website without having to solve a CAPTCHA.
    \item No
\end{Qlist}
}

\Qitem{ \Qq{What is your preference from the given options below?}
\begin{Qlist}
    \item It should not take me more than few seconds to solve a CAPTCHA.
    \item It should not take me more than one attempt to solve a CAPTCHA.
    \item The security aspect of a CAPTCHA should be the priority, and I do not care about the time spent or attempts taken to solve it.
    \item CAPTCHAs should be replaced by a new approach. 
\end{Qlist}
}

\Qitem{ \Qq{Have you ever run into accessibility problems while solving a CAPTCHA?}
\begin{Qlist}
    \item Yes (Please explain) \Qline{3cm}
    \item No
\end{Qlist}
}

\Qitem{ \Qq{Do you think CAPTCHA does a good job at differentiating bots from humans?}
\begin{Qlist}
    \item Yes, it eliminates the bots and their activities very well.
    \item Partially, it does a decent job at eliminating bots.
    \item No, I am fed up with bots.
\end{Qlist}
}

\Qitem{ \Qq{With the advancing technology, all CAPTCHA types today can be breached with a good accuracy?}
\begin{Qlist}
    \item Yes, there is no CAPTCHA type that cannot be breached.
    \item No, CAPTCHAs keep evolving and their security has evolved as well.
    \item I am not sure.
\end{Qlist}
}

\Qitem{ \Qq{What is your stance on CAPTCHA's security and usability aspects?}
\begin{Qlist}
    \item They are secure, but usability is a concern.
    \item They are usable, but security is a concern.
    \item They are secure as well as usable.
    \item Neither secure nor usable. 
\end{Qlist}
}

\end{subappendices}

%
%
%
\bibliographystyle{splncs04}
\bibliography{mybibliography}

\end{document}